\documentclass[twocolumn,showpacs,showkeys,preprintnumbers,amsmath,amssymb]{revtex4}

\usepackage{xcolor}
\usepackage{amsmath}
\usepackage[active]{srcltx}
\usepackage{graphicx}
\usepackage{dcolumn}
\usepackage{bm}
\usepackage{epsfig}
\usepackage{epstopdf}

\begin{document}

\preprint{IST/UL.2015-M J Pinheiro}

\title[Nuclear reactions]{Nuclear interaction modeled with a simple piston-gas model}

\author{Mario J. Pinheiro}
\email{mpinheiro@tecnico.ulisboa.pt}

\affiliation{Department of Physics, Instituto Superior T\'{e}cnico - IST, Universidade de Lisboa - UL, Av. Rovisco Pais, \&
1049-001 Lisboa, Portugal\\
Phone: 351.1.21.841.93.22\\
Fax: 351.1.21.846.44.55}

\homepage{http://mjpinheiro.weebly.com/}

\thanks{}

\date{\today}

\begin{abstract}
A simple one-dimensional gas-piston kinetic model gives the interaction potential between two colliding heavy ions. In the frame of the classical, thermodynamical approach, the colliding heavy ions are not submitted to friction, but produces an irreversible phenomena with cause at the difference of pressure $p$ ``felt" by the nucleon gas when ions collide with the target when compared with the pressure that nuclear matter exert on their boundaries when in thermodynamical equilibrium, and offers a straightforward way to calculate interacting potentials.
\end{abstract}



\maketitle

\section{Introduction}

The numerical calculations implicated in the study of nuclear reactions between heavy ions have an extreme complexity. Some times it is advisable to build simple models that allow to obtain concrete images of the phenomena based on classical conceptions and made useful predictions. In this perspective, D. H. E. Gross~\cite{Gross_1975} proposed the paradigmatic piston-gas model. However, our interpretation differs, not sustaining a friction process.

We propose in this Letter a classical, thermodynamical  approach to determine the interaction potential to which two colliding heavy ions (with mass number $A$ above 40) are submitted, and its dependency relatively to their nuclear temperature $T$. The high number of nucleons present inside each nucleus allows a macroscopic treatment. Also, the high value of the relative momentum $(\hbar k)$ with $k\approx 10-40$ fm$^{-1}$ ensures that the reduced wavelength to be $\lambdabar =1/k \ll 1$ fm, which implies that the wave packet associated to each ion is much smaller that the ion radius ($\approx 1.2 A^{1/3}$ fm).

The collision reactions are irreversible and can be treated as a dissipative phenomena due to ``friction" that are supposed to be occurring at the moment of impact. However, we introduce here a different perspective, based on previous work done by R. de Abreu et al.~\cite{Abreu_1,Abreu_2}, consisting in the following assumption: there is no friction, but an irreversible phenomena with cause at the difference of pressure $p'$ ``felt" by the nucleon gas when ions collide with a target when compared with the pressure that nuclear matter exert in their boundaries when in thermodynamical equilibrium $p$:
\begin{equation}\label{eq1}
p'=p \left( 1-\frac{2 \dot{r}}{u} \right),
\end{equation}
with $\dot{r}$ and $u$ denoting, resp., the speeds of the piston and of the gas particles, supposed to collide in chaotical manner with the piston and in (relatively) great number.

It is known that the variation of energy due to the piston is given by
\begin{equation}\label{eq2}
dE=-dW_{diss}=p' dV=p\left(1-\frac{2\dot{r}}{u}  \right) Adr.
\end{equation}
Notice that we are treating here the repulsive region of the potential when the ions are compressing each other, and distant from the repulsive core, $r \gtrsim a$, otherwise we must consider the repulsive Coulomb potential as well.
As a first approximation, we assume $p$ independent from $r$, and we can immediately integrate Eq.~\ref{eq2} from $r=a$, with $r$ denoting a minimal approaching distance and $r$ a given distance from the center of the nucleus:
\begin{equation}\label{eq3}
E=\varphi(r)=A \int_{a}^{r} \left( 1-\frac{2\dot{r}}{u} \right)\frac{dr}{dt}dtB(T,\mu,m_N),
\end{equation}
where $A$ is the area of contact between the two nucleus and $B(T,\mu,m_N)$ is the pressure exerted by the nuclear matter, dependent on the gas temperature $T$, chemical potential $\mu$, and the nucleons mass $m_N$. We will assume that the gas of nucleons is a Fermi gas without interaction between its particles. The chemical potential is given by (see, e.g., Ref.~\cite{Landau})
\begin{equation}\label{eq4}
\Omega=-\frac{2}{3}\frac{gVm_N^{3/2}}{\sqrt{2}\pi^2 \hbar^3} \int_0^{\infty} \frac{\sqrt{\varepsilon} d\varepsilon}{e^{\frac{\varepsilon-\mu}{T}}+1}
\end{equation}
It can be shown that
\begin{equation}\label{eq5}
\frac{\partial \Omega}{\partial T}=-\frac{VTg\sqrt{2\mu}m_N^{3/2}}{6 \hbar^3}
\end{equation}
and the entropy is ($S=-(\partial \Omega/\partial T)_{V, \mu}$)
\begin{equation}\label{eq6}
S=VT\frac{g\sqrt{2\mu} m_N^{3/2}}{6\hbar^3}.
\end{equation}
The pressure is straightforwardly obtained:
\begin{equation}\label{eq7}
p=p(T=0) + \frac{T^2 g \sqrt{2 \mu} m_N^{3/2}}{12 \hbar^3},
\end{equation}
and it can be rewritten using Eq.~\ref{eq6}:
\begin{equation}\label{eq7a}
p=p(T=0) + \frac{T}{2}\frac{S}{V}.
\end{equation}
We assume that $p(T=0) \sim 0$. Eq.~\ref{eq7a} shows that the resulting compression due to the heavy ion collision is equivalent to an amount of work $TS/2$, and generation of entropy due to the new arrangement of particles. From the above, we obtain the interacting potential
\begin{equation}\label{eq8}
\varphi(r)=-A\frac{gT^2 \sqrt{2 \mu}m_N^{3/2}}{12 \hbar^3} \int_0^r \left(1-\frac{2 \dot{r}}{u}  \right)\frac{dr}{dt}dt.
\end{equation}
Keeping with the analogy between the system piston-gas and the collision between heavy ions, we assume that the the energy of the incident ion is given by
\begin{equation}\label{eq9}
\varepsilon=\frac{1}{2}M\dot{r}^2 + \varphi(r).
\end{equation}
It results
\begin{equation}\label{eq10}
\varphi(r)=-A B(T,\mu, m_N) \int_a^r \left( 1-\frac{2}{u}\sqrt{\frac{2}{M}}\sqrt{\varepsilon-\varphi(r')} \right) dr',
\end{equation}
with
\begin{equation}\label{eq10a}
B(T,\mu,m_N) \equiv \frac{gT^2 \sqrt{2 \mu}m_N^{3/2}}{12 \hbar^3}.
\end{equation}
We can go a step further to obtain
\begin{equation}\label{eq11}
\varphi(r)=-AB(r-a)+\frac{2}{u}\sqrt{\frac{2}{M}}AB \int_a^r \sqrt{\varepsilon - \varphi(r')}dr'
\end{equation}
where $M$ denotes the mass of the heavy ion and $a$ is the distance of minimal approximation, corresponding to the moment when the ions collide inelastically, dissipating energy and ejecting nucleons in the process.

\begin{figure}
  \centering
  \includegraphics[width=3.40 in]{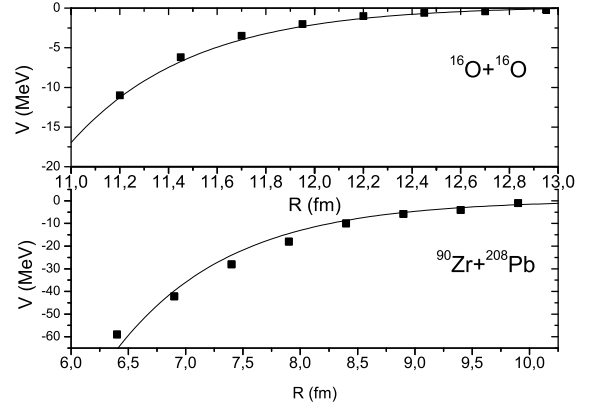}\\
  \caption{Ion-ion potential for reaction (a) ${}^{16}O+{}^{16}O$ and (b) ${}^{90} Zr + {}^{208} Pb$. The touching point distance of two ions is $a=6,9$ fm, $\phi(a)=-5.77$ MeV, $C=0.26$. The square dots are data representing the calculations obtained in the Hartree-Fock-Bogoliubov approximation with SkM$^{*}$ parameter set of the Skyrme force~\cite{Denisov_2002}.}\label{Fig1}
\end{figure}

The mathematical resolution of Eq.~\ref{eq11} is complex since we don't have a previous idea of what is the form of the interaction potential $\varphi (r)$ (in fact, it is what we want to know). However, we may assume that the total energy of the heavy ion is much higher that the interaction potential in the range considered (until the minimal distance), $\varepsilon \gg \varphi (r)$ (relativistic heavy-ions collisions), in which case we obtain a more simplified integral equation:
\begin{equation}\label{eq12}
\varphi(r)=f(r)+\lambda \int_a^r \varphi(r)dr.
\end{equation}
Here,
\begin{equation}\label{eq13}
f(r) \equiv - [A B(T, \mu, m_N) + \sqrt{\varepsilon}](r-a),
\end{equation}
and
\begin{equation}\label{eq14}
\lambda \equiv -\frac{1}{u}\sqrt{\frac{2}{M\varepsilon}} AB(T,\mu, m_N).
\end{equation}

In the relativistic limit, the mathematical solution of Eq.~\ref{eq12} is given by
\begin{equation}\label{eq15}
\varphi(r)=\frac{C}{\lambda}+\varphi(a) e^{- \mid \lambda | (r-a)}.
\end{equation}
$C$ is a constant of integration that appears after converting Eq.~\ref{eq12} to a first-order differential equation.
In Fig.~\ref{Fig1} it is shown the calculations using Eq.~\ref{eq15} for the ion-ion potential for reaction ${}^{16}O+{}^{16}O$. The touching point distance of two ions is $a=6.9$ fm for the set of three reactions shown in Table~\ref{Table1} despite the different number of nucleons in each nucleus, $\phi(a)=-7.0$ MeV, $C=0.26$ Mev$/$fm, and $\mid \lambda \mid = 1$ fm$^{-1}$. Notice that $r_c =1/\mid \lambda \mid =1$ fm is the critical distance in remarkable agreement with the expected value~\cite{Mosel_1974}. In the Woods-Saxon spherical potential describing single-particle levels, $r_c$ represents the surface thickness (e.g., Ref.~\cite{Greiner_1996}).

The comparison is made with several calculations obtained in the Hartree-Fock-Bogoliuobov approximation with SkM$^{*}$ parameter set of the Skyrme force~\cite{Denisov_2002} and our theoretical approximation of Eq.~\ref{eq15} with parameters shown in Table~\ref{Table1}. 

The advantage of the method proposed in this Letter stays in the possibility to obtain an analytical form for the interaction potential, more easy, fast and precise to use. The simple physical model also points out that what occurs between the colliding heavy ions is not a friction process, but simply an irreversible process due to the difference of pressures felt by the nucleon gas before and during the collisional process.

\begin{figure}
  \centering
  \includegraphics[width=3.6 in]{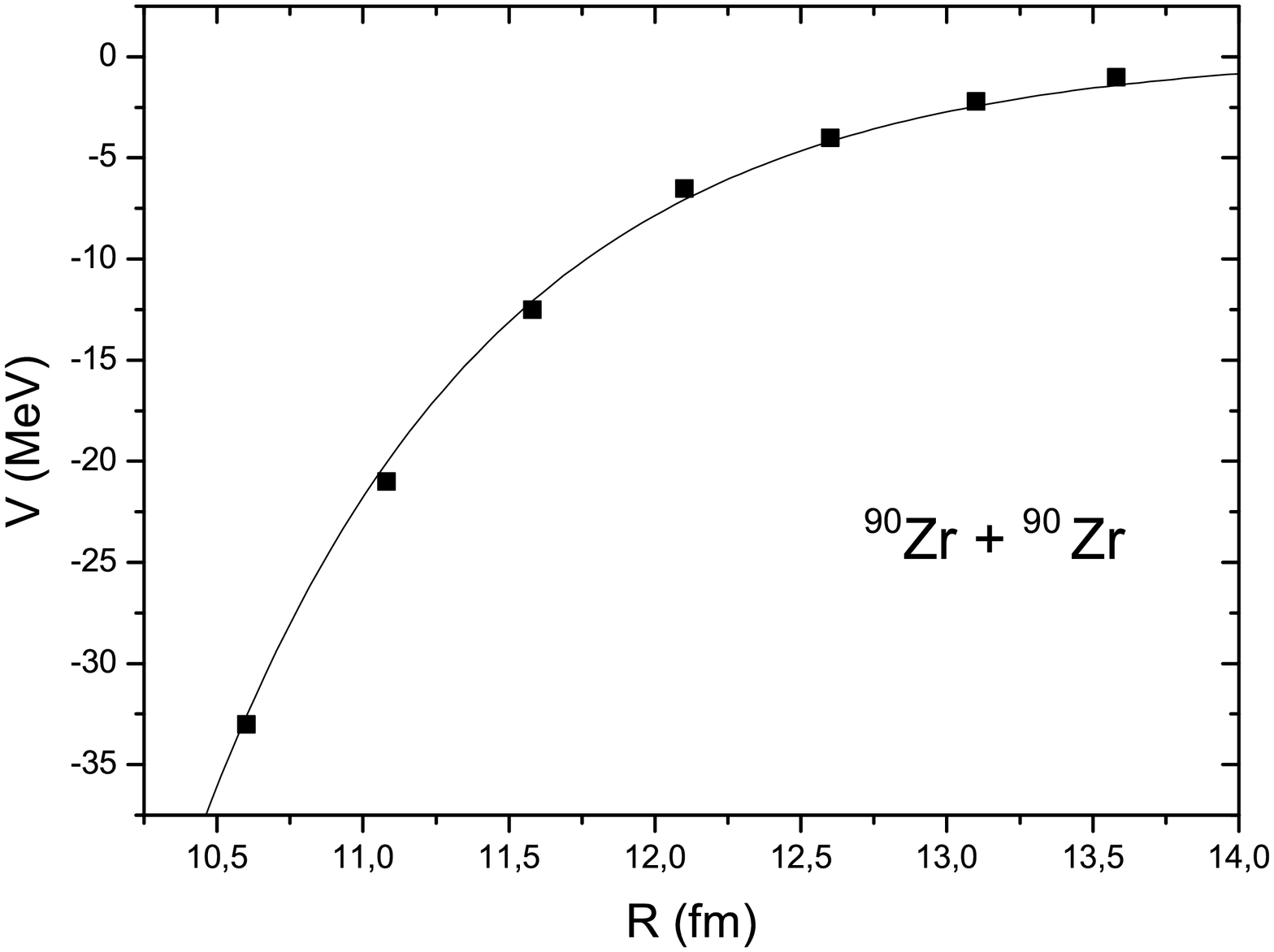}\\
  \caption{Ion-ion potential for ${}^{90} Zr + {}^{90} Zr$. The square dots are data representing the calculations obtained in the Hartree-Fock-Bogoliubov approximation with SkM$^{*}$ parameter set of the Skyrme force~\cite{Denisov_2002}.}\label{Fig2}
\end{figure}

\begin{table}
  \centering
  \caption{Parameters for the interaction potential in nuclear heavy ions collisions. Potentials best fitted for $\lambda=1$ fm$^{-1}$.}\label{Table1}
\begin{tabular}{c c c c}
  \hline
  Nuclear reaction          & $a$ (fm)              & $\phi(a)$ (MeV)& C (MeV$/$fm)\\ [0.25 cm]
  ${}^{16}O+{}^{16}O$       & 6.9                   & -7.0           & 0.26\\ [0.25 cm]
  ${}^{90} Zr + {}^{90} Zr$ & 6.9                   & -1330.0        & 0.26 \\[0.25 cm]
  ${}^{90} Zr + {}^{208} Pb$& 6.9                   & -40.0          & 0.26 \\ [0.25 cm]
  \hline
\end{tabular}
\end{table}

\section{Conclusion}

It was obtained the interaction potential between two heavy ions in the framework of a simple, classical, one-dimensional piston-gas model, where there is no friction, but an irreversible phenomena with cause at the difference of pressure $p'$ ``felt" by the nucleon gas when ions collide with a target when compared with the pressure that nuclear matter exert in their boundaries when in thermodynamical equilibrium $p$. Our parametrized potentials are in good agreement with other theoretical potentials, particularly with the Hartree-Fock-Bogoliuobov approximation with SkM$^{*}$ parameter set of the Skyrme force, and with experimental data.

\begin{acknowledgments}
The author gratefully acknowledge partial financial support by the Funda\c{c}\~{a}o para a Ci\^{e}ncia e Tecnologia under contract ref. SFRH/BSAB/1420/2014.
\end{acknowledgments}

\bibliographystyle{apsrev}
\end{document}